\documentclass[sn-nature,iicol]{sn-jnl}

\usepackage{graphicx}%
\usepackage{multirow}%
\usepackage{amsmath,amssymb,amsfonts}%
\usepackage{amsthm}%
\usepackage{mathrsfs}%
\usepackage[title]{appendix}%
\usepackage{xcolor}%
\usepackage{textcomp}%
\usepackage{manyfoot}%
\usepackage{booktabs}%
\usepackage{algorithm}%
\usepackage{algorithmicx}%
\usepackage{algpseudocode}%
\usepackage{listings}%

\begin{document}

\title[Re-investigation of Moment Direction in Kitaev Material $\alpha$-RuCl$_{3}$]{Re-investigation of Moment Direction in a Kitaev Material $\alpha$-RuCl$_{3}$}

\author[1,2]{\fnm{Subin} \sur{Kim}}\email{skim6@bnl.gov}

\author[1]{\fnm{Ezekiel J.} \sur{Horsley}}\email{ezekiel.horsley@utoronto.ca}

\author[2]{\fnm{Christie S.} \sur{Nelson}}\email{csnelson@bnl.gov}

\author[3]{\fnm{Jacob P. C.} \sur{Ruff}}\email{jpr243@cornell.edu}

\author*[1]{\fnm{Young-June} \sur{Kim}}\email{youngjune.kim@utoronto.ca}

\affil[1]{\orgdiv{Department of Physics}, \orgname{University of Toronto}, \city{Toronto}, \postcode{M5S 1A7}, \state{Ontario}, \country{Canada}}

\affil[2]{\orgdiv{National Synchrotron Light Source II}, \orgname{Brookhaven National Laboratory}, \city{Upton}, \postcode{11973}, \state{NY}, \country{USA}}

\affil[3]{\orgdiv{Cornell High Energy Synchrotron Source}, \orgname{Cornell University}, \city{Ithaca}, \postcode{14853}, \state{NY}, \country{USA}}

\abstract{We report X-ray diffraction and resonant elastic X-ray scattering (REXS) studies on two $\alpha$-RuCl$_{3}$ crystals with distinct magnetic transition temperatures: T$_{N}$=7.3K and 6.5K. We find that the sample with T$_{N}$=6.5K exhibits a high degree of structural twinning at low temperature, whereas the T$_{N}$=7.3K sample primarily comprises a single domain of R$\bar{3}$. Notwithstanding, both samples exhibit an identical zigzag magnetic structure, with magnetic moments pointing away from the honeycomb plane by $\alpha=31(2)^{\circ}$. We argue that the identical ordered moment directions in these samples suggest that the intralayer magnetic Hamiltonian remains mostly unchanged regardless of T$_{N}$.}

\maketitle

\section{Introduction}

In recent years, $\alpha$-RuCl$_{3}$ has been extensively investigated as a potential candidate for a Kitaev quantum spin liquid \cite{Winter2017,Takagi2019,Motome2020,Kim2022,Plumb2014,sears15,banerjee16,Banerjee2018,Do2017,Balz_2021,sandilands15,zheng2017}. The intriguing interplay of frustrating, bond-dependent interactions arising from spin-orbit coupling offers a promising platform for having a quantum spin liquid \cite{jackeli2009}. While the system exhibits magnetic ordering below T$_{N}\approx$7K, it can be effectively suppressed by applying a moderate magnetic field parallel to the honeycomb plane \cite{Sears2015,Johnson2015,banerjee16,Balz_2021,Sears2017}. This field-induced quantum phase has been the subject of many studies \cite{baek2017,Banerjee2018,Kasahara2018,Ponomaryov2020,leahy2017,Modic2021,Kasahara2018prl,yokoi2021,yamashita2020,Li2023,Gass2020}, including the observation of the half-quantized thermal Hall effect, which is regarded as a compelling piece of evidence of a potential quantum spin liquid phase \cite{Kasahara2018}. However, this experimental observation remains contentious due to limited reproducibility \cite{czajka2021,Lefrancois2022}. The thermal Hall effect displays a significant dependence on the sample quality and composition, and the quantization was only observed in select samples \cite{Kasahara2018,czajka2021,Lefrancois2022,tanaka2022,Kasahara2022,bruin2022,Yamashita2020prb}. 

\begin{figure*} [ht]
    \includegraphics[width=1\textwidth]{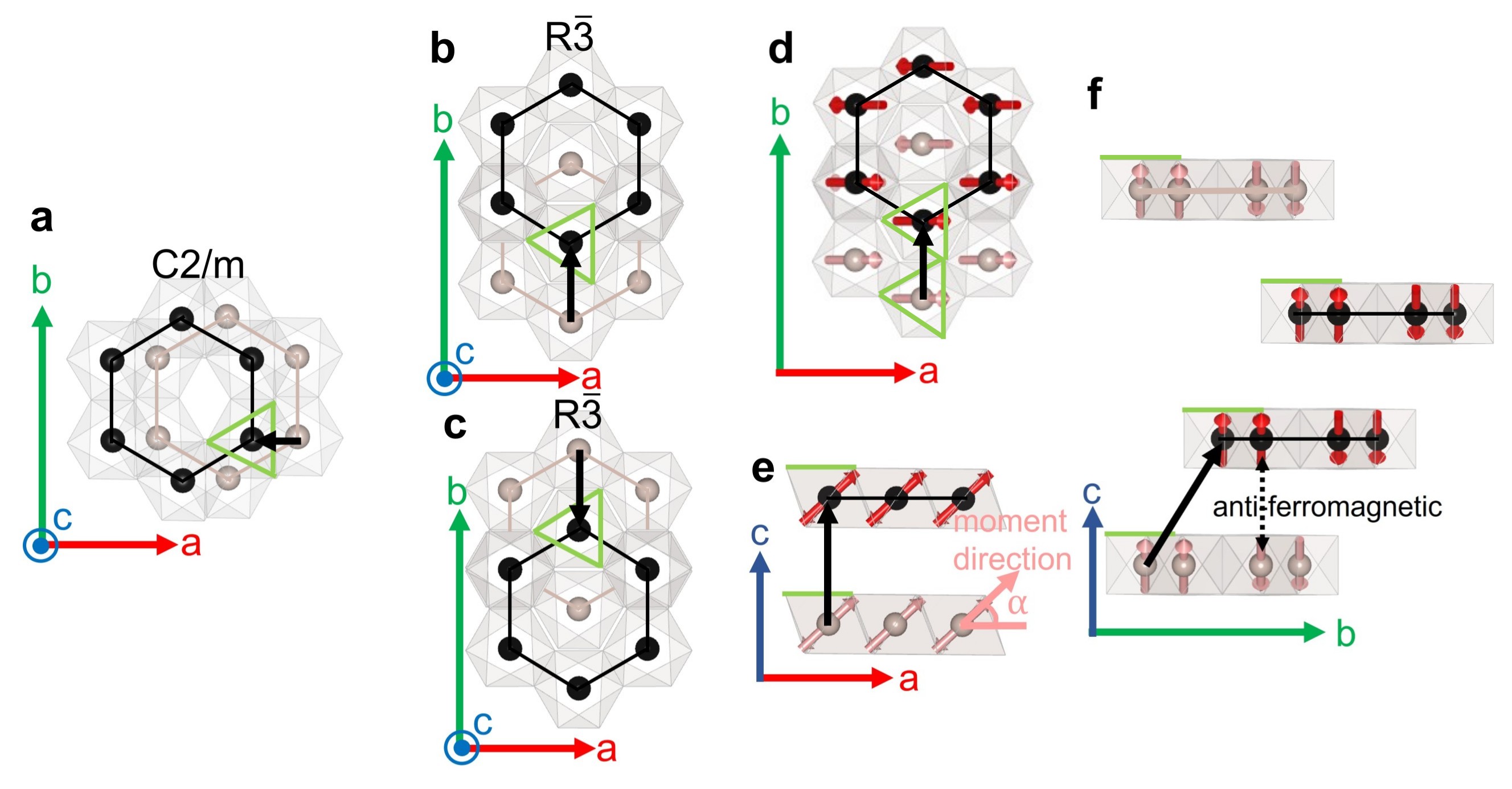}
    \caption{\textbf{Lattice and Magnetic Structure of $\alpha$-RuCl$_{3}$}. \textbf{a} High-temperature C2/m structure. Adjacent layers are stacked with a relative shift along $\vec{a}$. We use the pseudo-orthorhombic coordinate system as described in the text. \textbf{b}-\textbf{c} Low-temperature R$\bar{3}$ structure of two twin domains. Adjacent layers are relatively shifted along $\vec{b}$ but in opposite directions. \textbf{d}-\textbf{f} Zigzag antiferromagnetic structure of $\alpha$-RuCl$_{3}$ projected onto different planes. The magnetic moment vector lies within the a-c plane, canted away from the honeycomb plane by angle $\alpha$ as shown in \textbf{e}. The magnetic structure is three-layer-periodic perpendicular to the honeycomb plane which can be explained by nearest neighbour antiferromagnetic coupling shown in \textbf{f}.}
    \label{fig:structure+mstructure}
\end{figure*}

The quality of the crystals varies among samples based on the amount of defects present in the crystal structure. $\alpha$-RuCl$_{3}$ features a honeycomb structure with weak coupling between its layers through van der Waals forces, making it susceptible to structural defects known as stacking faults (see Fig.~\ref{fig:structure+mstructure}). This inherent predisposition to stacking sequence disorder results in diverse structural arrangements among crystals, leading to the proposition of multiple crystal structures for this material \cite{johnson15,banerjee16,cao16,bruin2022,stahl2022}. Careful studies using high-quality $\alpha$-RuCl$_{3}$ crystals indicate the crystal structure symmetry is C2/m at room temperature. However, this structure is disrupted below the structural phase transition around 150K. The exact identification of the low-temperature structure has been controversial, but recent studies converge on $R\bar{3}$ symmetry as the low-temperature structure (see Fig.~\ref{fig:structure+mstructure}\textbf{a}) \cite{park2016,Mu2022,Zhang2023,Kim2023}.

The magnetic structure of $\alpha$-RuCl$_{3}$ is sample-dependent as well. Early neutron diffraction studies on a single crystal by Sears et al. observed magnetic Bragg peaks with three-layer periodicity \cite{sears15}. In contrast, neutron diffraction studies on a powder sample by Johnson et al. observed magnetic Bragg peaks with two-layer periodicity instead \cite{johnson15}. Subsequently, Banerjee et al. showed that the three-layer and two-layer periodic magnetic structures have distinct magnetic transition temperatures, T$_{N}$=7~K and T$_{N}$=14~K, respectively \cite{banerjee16}. The magnetic structure characterized by a two-layer periodicity is associated with samples with a substantial number of stacking faults. This correlation is emphasized by the presence of prominent diffuse scattering, notably observed in samples with T$_{N}$=14~K \cite{Kim2022}. This correlation is reinforced by the difficulty in reconciling a two-layer magnetic periodicity with the low-temperature R$\bar{3}$ structure \cite{cao16}. Johnson et al. explained this two-layer periodicity by invoking the high-temperature C2/m structure instead \cite{Johnson2015}, suggesting that the C2/m structure persists down to low temperature in these samples with T$_{N}$=14~K. On the other hand, the three-layer magnetic structure with T$_{N}$=7~K is typically observed for samples without strong diffuse scattering \cite{Kim2022}, and it can be explained exclusively using the R$\bar{3}$ structure (see Fig.~\ref{fig:structure+mstructure}) \cite{Balz_2021}. Therefore, it is generally accepted that crystals with a single magnetic transition around T$_{N}$=7K are of high quality.

However, in addition to this large variation in T$_{N}$, recent reports showed that a smaller variation in T$_{N}$ is found even among high-quality crystals, with values ranging from 6.5K to 7.5K \cite{Sears2017,Do2017,Kasahara2022,bruin2022_oscillation,Zhang2023}. While these studies suggest that samples with higher T$_{N}$ tend to be of better quality, the precise origin of this sample-dependent variation in T$_{N}$ remains unresolved \cite{Zhang2023}. It is this small variation that is the main subject of this paper. Specifically, we examine whether the magnetic structure and the ordered moment direction are different between these samples. We carried out a resonant elastic X-ray scattering (REXS) study of two samples with different magnetic transition temperatures of T$_{N}$=6.5K and T$_{N}$=7.3K. We confirm the zigzag antiferromagnetic arrangement with three-layer periodicity in both samples (See Fig.~\ref{fig:structure+mstructure}\textbf{d}-\textbf{f}). In particular, a study of  the azimuthal angle dependence shows that the out-of-plane canting angle $\alpha$ remains unchanged between the two samples (see Fig.~\ref{fig:structure+mstructure}\textbf{e}). In addition, the observed canting angle is consistent with that previously observed for a sample with T$_{N}$=12K and two-layer-periodicity \cite{Sears2020}. The fact that no discernible differences in the moment direction were observed across these samples suggests that the magnetic Hamiltonian remains unchanged regardless of the sample quality. The only distinction we find between the two samples we studied pertains to their structural domain populations. The sample exhibiting T$_{N}$=7.3K is predominantly comprised of a single domain of the R$\bar{3}$ structure, while the T$_{N}$=6.5K sample showed highly twinned R$\bar{3}$ structures.

\section{Results}
\subsection{Overview of the Crystals}\label{Overview}

Minor variations in the transition temperature was confirmed between 5 $\alpha$-RuCl$_{3}$ crystals, for which the T$_{N}$ values varied from 6.5K to 7.3K (see Supplementary Information). The specific heat and the magnetic susceptibility measurements suggest a distinction in the crystal quality between samples with T$_{N}$=6.5(1)~K and samples with T$_{N}$=7.3(1)~K. Here, we choose one sample from each group: S1 with T$_{N}$=7.3K and S3 with T$_{N}$=6.5K, and investigated their crystal and magnetic structures.

\subsection{Structural Characterization}\label{SC}
\begin{figure*} [ht]
    \centering
    \includegraphics[width=0.7\textwidth]{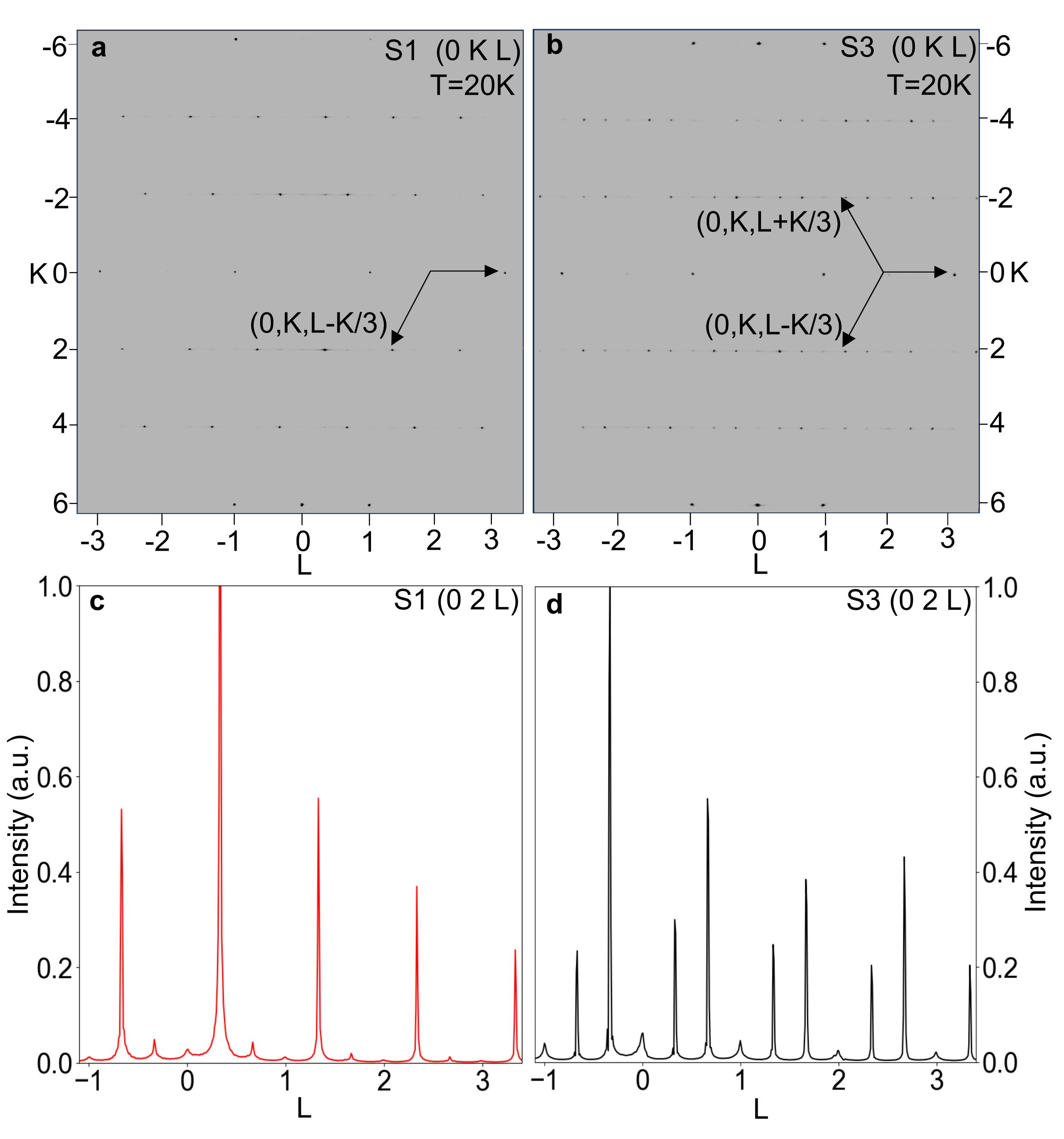}
    \caption{\textbf{X-ray diffraction results} Maps of structural peaks at T=20K in the reciprocal 0KL plane of \textbf{a} S1 and \textbf{b} S3. In the case of S1, the Bragg peaks are discernible at (0,K,L-K/3). For S3, additional Bragg peaks are evident at (0,K,L+K/3) together with (0,K,L+K/3). Each of these Bragg peak types: (0,K,L-K/3) and (0,K,L+K/3) corresponds to the two possible twin domain structures shown in Fig.~\ref{fig:structure+mstructure}\textbf{b} and \textbf{c}, respectively. \textbf{c}-\textbf{d} (0,2,L) line scan of S1 and S3, respectively. Predominant peaks are observable at (0,2,L-2/3) in the case of S1, while additional peaks manifest at (0,2,L+2/3) for S3. The minor peak centered at (0,2,L), which displays a broader width compared to other Bragg peaks, is derived from the high-temperature C2/m structure depicted in Fig.~\ref{fig:structure+mstructure}\textbf{a}.}
    \label{fig:xray}
\end{figure*}

Both samples undergo a first-order structural transition from a monoclinic C2/m structure above the structural transition temperature, T$_{s}\approx$150K, to a rhombohedral R$\bar{3}$ structure below it, as depicted in Fig.~\ref{fig:structure+mstructure}.
To describe the two structures using the same coordinate system, we adopt a pseudo-orthorhombic notation as shown in Fig.~\ref{fig:structure+mstructure}. The $\vec{a}$ and $\vec{b}$ vectors are identical to the monoclinic structure, and describe two unique high-symmetry directions within the honeycomb structure. The $\vec{c}$ vector points perpendicular to the honeycomb plane and describes the separation vector between adjacent layers. Note that this $\vec{c}$ is not a lattice vector for either the C2/m or the R$\bar{3}$ structure. In the C2/m structure, each honeycomb layer is stacked on top of another with a shift by $-\vec{a}$/3. On the other hand, the layers are shifted by $\vec{b}$/3 in the R$\bar{3}$ structure. The primary distinction between these two structures lies in their stacking direction, and this can be differentiated using single crystal X-ray diffraction.

Fig.~\ref{fig:xray}\textbf{a}-\textbf{b} show X-ray diffraction maps, comparing the reciprocal (0,K,L) plane diffraction patterns of S1 and S3 at 20K, well below the structural transition. In the case of S1, Bragg peaks are found at (0,K,L-K/3), where K is an even integer and L is an integer. However, for S3, additional Bragg peaks are also visible at (0,K,L+K/3), together with (0,K,L-K/3). Both (0,K,L-K/3) and (0,K,L+K/3) types of Bragg peaks arise from the R$\bar{3}$ structure but they correspond to different structural twin domains, as shown in Fig.~\ref{fig:structure+mstructure}\textbf{b}-\textbf{c}. The translation vectors $\pm$$\vec{b}$/3+$\vec{c}$ for each twin domain give rise to Bragg peaks at (0,K,L$\mp$K/3), respectively. Therefore, we can conclude that S1 mainly comprises a single domain, while S3 has a significant mixture of twin domains. This distinction becomes clearer when examining L scans along (0,2,L), illustrated in Fig.~\ref{fig:xray}\textbf{c}-\textbf{d}. In S1, intense peaks are found at (0,2,L-2/3), with much smaller peaks present at (0,2,L+2/3). However, in the case of S3, significant intensity is present at both (0,2,L$\pm$2/3), indicative of substantial twinning at a ratio of about 2:1. Additionally, weak peaks at (0,2,L) are present (more pronounced when compared to S1). These peaks indicate that a small fraction of the sample remains in the high-temperature C2/m structure. The origin of this is unclear, but it may be related to an incomplete transition from the high-temperature phase. Alternatively, the C2/m phase may act as a domain wall between two twin domains. The widths of these Bragg peaks are much broader than that of the R$\bar{3}$ counterparts in which the domain size is roughly 20 layers. The extensive twinning combined with the presence of the high-temperature structure in S3 results in a prominent diffuse rod of intensity along $\vec{c}$ (see Fig.~\ref{fig:xray}\textbf{b}) in comparison to S1. This observation agrees with Zhang et al. who reported that samples with lower T$_{N}$ exhibit larger diffuse scattering \cite{Zhang2023}.

\subsection{Magnetic Structure}\label{MS}
\begin{figure*} [ht]
    \centering
    \includegraphics[width=1\textwidth]{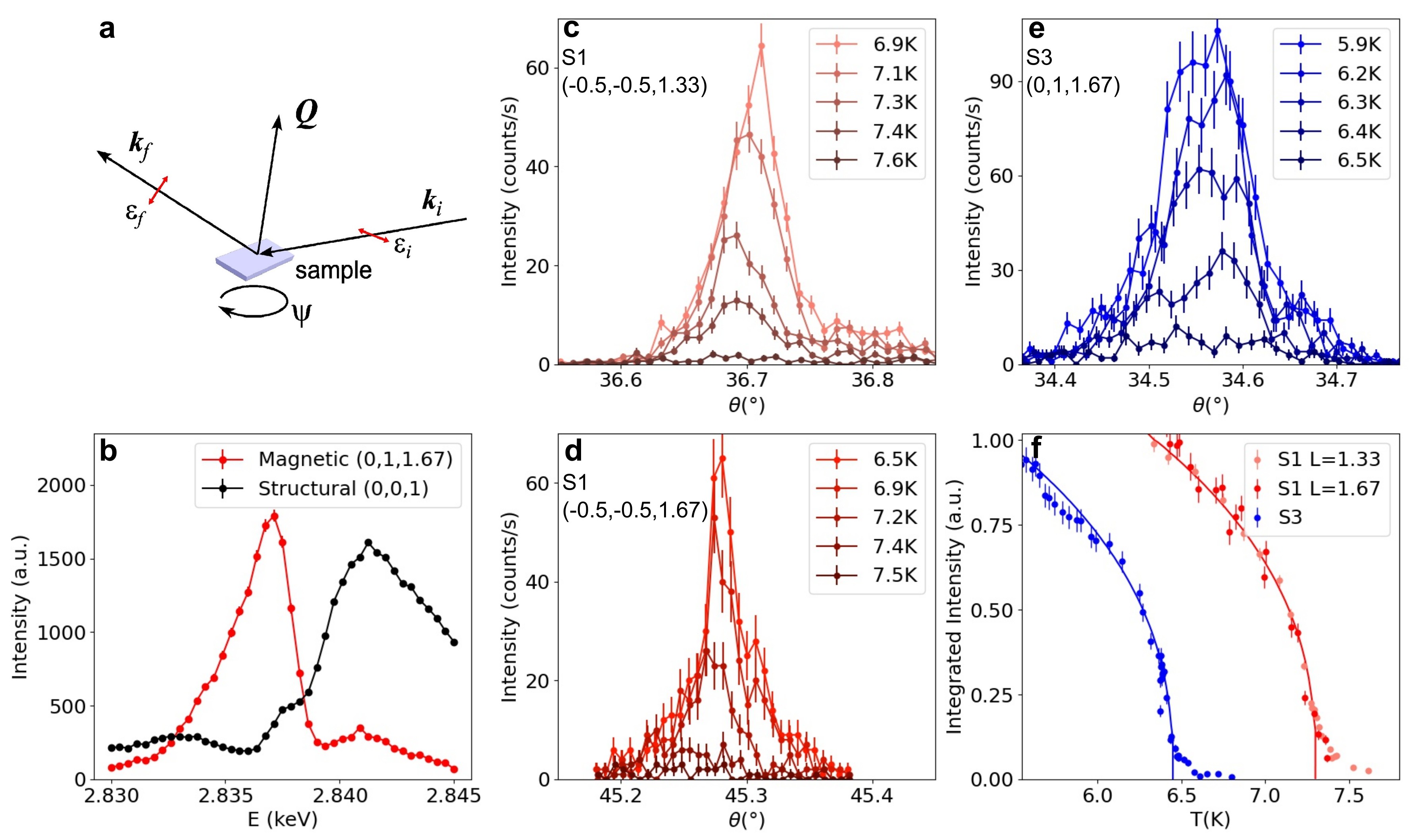}
    \caption{\textbf{Resonant elastic X-ray scattering - temperature dependence} \textbf{a} Schematic diagram of the scattering geometry of the REXS experiment. $\psi$ represents azimuthal angle when the sample is rotated with respect to scattering vector ${\bf Q}={\bf k}_{f}-{\bf k}_{i}$. \textbf{b} Photon energy dependence of the structural (0,0,1) and magnetic (0,1,1.67) Bragg peaks intensities. The intensity of the magnetic Bragg peak resonates at E=2.837keV while the intensity of the structural Bragg peak is suppressed due to absorption. \textbf{c}-\textbf{d} Rocking curves of magnetic Bragg peak ${\bf Q}_{m1}=(-0.5,-0.5,1.33)$ and ${\bf Q}_{m1}^{\prime}=(-0.5,-0.5,1.67)$ respectively in S1, at various temperatures. Both magnetic Bragg peaks are suppressed upon heating, vanishing above the magnetic transition temperature of T$_{N}$=7.3~K. \textbf{e} Sample rocking curve of magnetic Bragg peak ${\bf Q}_{m2}^{\prime}=(0,1,1.67)$ in S3 at varying temperatures. The magnetic Bragg peak vanished above the magnetic transition temperature at T$_{N}$=6.5~K. \textbf{f} Integrated intensity of the rocking curves \textbf{c}-\textbf{e} as a function of temperature. The lines are fitted to $\sim (1-T/T_N)^{0.4}$ to extract the transition temperatures. A clear difference in T$_{N}$ can be seen between the two samples.}
    \label{fig:schematic+temperature}
\end{figure*}

The magnetic structures of S1 and S3 were studied using REXS. In both samples, the magnetic Bragg peaks were observed using the $\sigma$-$\pi'$ polarization channel for which the incident beam is polarized perpendicular to the scattering plane ($\sigma$) and the scattered beam is polarized parallel to the scattering plane ($\pi$'), as illustrated in Fig.~\ref{fig:schematic+temperature}\textbf{a}. The magnetic Bragg peak intensities show resonant enhancement at the L$_{3}$ edge of Ruthenium (2.837keV), as shown in Fig.~\ref{fig:schematic+temperature}\textbf{b}. In the case of S1, magnetic Bragg peaks were observed at ${\bf Q}_{m1} = (-0.5, -0.5, 1.33)$ and ${\bf Q}_{m1}^{\prime} = (-0.5, -0.5, 1.67)$. Here, the primed notation indicates an equivalent {\bf Q}-position in the other rhombohedral twin domain (see Fig. 1b and 1c), as discussed below. For S3, magnetic Bragg peaks were also observed at ${\bf Q}_{m2} = (0, 1, 1.33)$ and ${\bf Q}_{m2}^{\prime} = (0, 1, 1.67)$, in addition to ${\bf Q}_{m1}$ and ${\bf Q}_{m1}^{\prime}$. Note that ${\bf Q}_{m1}$ and  ${\bf Q}_{m2}$ as well as  ${\bf Q}_{m1}^{\prime}$ and ${\bf Q}_{m2}^{\prime}$ are equivalent under the three-fold rotational symmetry of the rhombohedral R$\bar{3}$ structure (see Supplementary Information).

Fig.~\ref{fig:schematic+temperature}\textbf{c}-\textbf{f} presents the temperature-dependence of magnetic Bragg peaks. Specifically, Fig.~\ref{fig:schematic+temperature}\textbf{c} and \textbf{d} demonstrate the temperature dependence of the sample rocking curve scans for ${\bf Q}_{m1}$ and ${\bf Q}_{m1}^{\prime}$ in S1, respectively. The temperature dependence of the rocking curve scans at ${\bf Q}_{m2}^{\prime}$ for S3 are shown in Fig.~\ref{fig:schematic+temperature}\textbf{e}. In each case, the magnetic Bragg peak intensity decreases with increasing temperature, confirming their magnetic origin. The temperature dependence of the integrated intensity is plotted in Fig.~\ref{fig:schematic+temperature}\textbf{f}, clearly illustrating the difference in the magnetic transition temperatures between the two samples. For S1, the magnetic Bragg peaks vanish around T$_{N}$=7.3(1)~K for both twin domains while in S3, the magnetic Bragg peak vanishes around T$_{N}$=6.5(1)~K. These transition temperatures agree well with the specific heat data. (see Supplementary Information) The magnetic Bragg peak intensity is proportional to the magnetic order parameter squared. The intensity data in Fig.~\ref{fig:schematic+temperature}\textbf{f} could be fitted to $\sim (1-T/T_N)^{2\beta}$. The critical exponent $\beta$ could not be determined with high accuracy due to large error bars. However, $\beta \approx 0.2$ describes the data fairly well, which indicates the two-dimensional nature of the phase transition. 

\begin{figure*} [ht]
    \centering
    \includegraphics[width=1\textwidth]{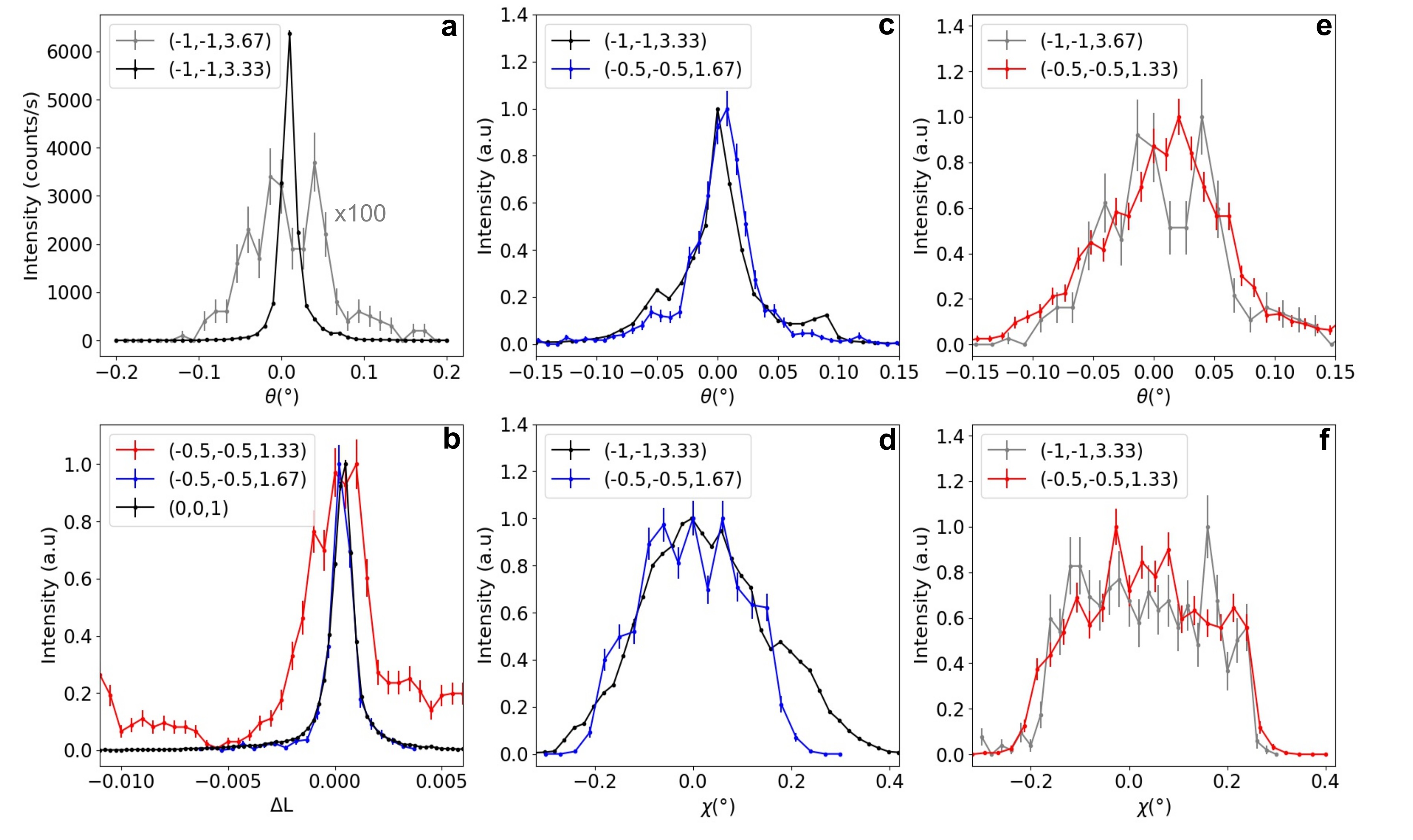}
    \caption{\textbf{Sample rocking curve scans comparison between structural and magnetic Bragg peaks} \textbf{a} Rocking curve comparison between two structural Bragg peak ${\bf Q}_{s1}$=(-1,-1,3.67) and ${\bf Q}'_{s1}$=(-1,-1,3.33) for S1. The intensity is multiplied 100 times for ${\bf Q}_{s1}$. \textbf{b} L scan comparison between magnetic Bragg peaks ${\bf Q}_{m1}$ and ${\bf Q}'_{m1}$ with structural Bragg peak (0,0,1). To facilitate this comparison, the scan is plotted against relative change in L, $\Delta$L. The width of magnetic Bragg peak ${\bf Q}'_{m1}$ is consistent with structural (0,0,1) while the width of ${\bf Q}_{m1}$ is observed to be much broader. \textbf{c}-\textbf{d} rocking curve scans along $\theta$ and $\chi$ respectively, comparing magnetic ${\bf Q}'_{m1}$ with structural ${\bf Q}'_{s1}$. \textbf{e}-\textbf{f} Rocking curves along $\theta$ and $\chi$ respectively, comparing magnetic ${\bf Q}_{m1}$ with structural ${\bf Q}_{s1}$.}
    \label{fig:rocking}
\end{figure*}

We first focus on S1 to examine its lattice and magnetic structure more carefully. Fig.~\ref{fig:rocking}\textbf{a} compares the scans of structural Bragg peaks ${\bf Q}_{s1}^{\prime}=(-1,-1,3.33)$ and ${\bf Q}_{s1}=(-1,-1,3.67)$, which are equivalent to (0,2,3.33) and (0,2,3.67), respectively. As discussed in the previous section, these peaks arise from different twin domains (corresponding to Fig.~1c and Fig.~1b, respectively), explaining large discrepancies in shape and intensity between them. In particular, the intensity differs by two orders of magnitudes, confirming that S1 comprises a predominantly single domain of R$\bar{3}$.  

In Fig.~\ref{fig:rocking}\textbf{b}, we present L scans for the magnetic Bragg peaks ${\bf Q}_{m1}$ and ${\bf Q}'_{m1}$, compared with the structural Bragg peak (0,0,1). A noticeable width difference is observed between ${\bf Q}_{m1}$ and ${\bf Q}'_{m1}$. While the width of the ${\bf Q}'_{m1}$ peak is resolution-limited and equals that of the structural Bragg peak (0,0,1), the width of the ${\bf Q}_{m1}$ peak is not resolution-limited, indicating a correlation length of about 250 layers. This difference in correlation lengths strongly suggests that they originate from different magnetic domains.

Furthermore, we can identify that these two magnetic peaks originate from the two structural twin domains. Fig.~\ref{fig:rocking}\textbf{c} and \textbf{d} show sample rocking curves along $\theta$ and $\chi$, respectively, for the magnetic ${\bf Q}'_{m1}$ peak, compared with the structural ${\bf Q}'_{s1}$ peak. Similarly, Fig.~\ref{fig:rocking}\textbf{e} and \textbf{f} display rocking curves along $\theta$ and $\chi$, respectively, for the magnetic ${\bf Q}_{m1}$ peak, compared with the structural ${\bf Q}_{s1}$ peak. Each pair of the magnetic and structural peaks agree well in their shape profiles which strongly suggest that they originate from the same domains. Since ${\bf Q}_{s1}$ and ${\bf Q}'_{s1}$ correspond to different structural twin domains of $R\bar{3}$, the observed ${\bf Q}_{m1}$ and ${\bf Q}'_{m1}$ peaks correspond to magnetic domains originating from different structural domains. See Supplementary Information for these peak positions. Note that each pair follows the same selection rule (0,K,L$\mp$K/3), respectively. The magnetic structure that best accounts for this observation is the three-layer periodic zigzag structure shown in Fig.~\ref{fig:structure+mstructure}. For S3, unravelling the magnetic structure proves challenging due to the presence of extensive twinning in both the structure and magnetic structure. Nonetheless, the intensity can still be well-explained using the magnetic structure provided in Fig.~\ref{fig:structure+mstructure}.

It is interesting to note that during the initial cool-down of S1, only the magnetic Bragg peak at ${\bf Q}'_{m1}$ was observed. However, subsequent cool-downs introduced an additional strong Bragg peak at ${\bf Q}_{m1}$ (see Supplementary Information). This suggests the possibility of a change in the magnetic domain configuration after thermal cycling. However, the situation seems to be much more complicated than a simple history dependence if one considers the relatively weak intensity of the corresponding structural peak at ${\bf Q}_{s1}$. Note that the structural peaks are observed using photons with the 3rd harmonic of the primary X-ray beam energy (E=8.511keV), while the magnetic peaks are probed using photons with the primary energy (E=2.837keV). The penetration depths for these beams are 30 µm and 1 µm, respectively. We speculate that the thermal cycling might affect the near-surface region of the sample disproportionally, giving rise to a proliferation of twin domain walls, and the corresponding change in the magnetic Bragg peak intensity.

\begin{figure*} [ht]
    \centering
    \includegraphics[width=1\textwidth]{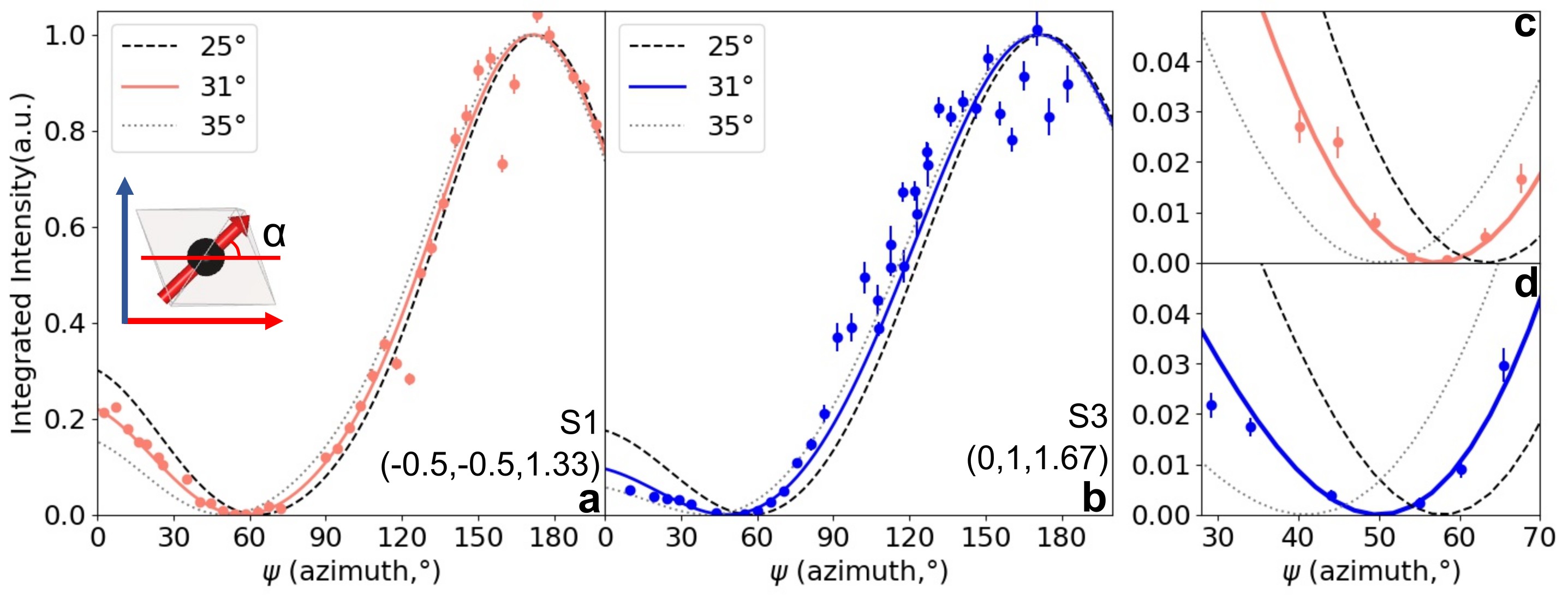}
    \caption{ \textbf{Azimuthal dependence of magnetic intensity}\textbf{a}-\textbf{b} Azimuthal dependence of integrated intensity of magnetic Bragg peaks of S1 and S3 respectively. In case of S1, azimuthal dependence of ${\bf Q}_{m1}$ is shown while in S3, azimuthal dependence of ${\bf Q}'_{m2}$ is shown instead. The lines are simulated azimuthal dependence for various canted moment angles, $\alpha$. The observed azimuthal dependence agrees with $\alpha$=31$^{\circ}$ \textbf{c}-\textbf{d} Azimuthal dependence close to zero intensity for S1 and S3. The vanishing angle $\psi_{c}$ agrees well with $\alpha$=31$^{\circ}$ for both S1 and S3.}
    \label{fig:azi}
\end{figure*}

\subsection{Magnetic Moment Direction}

In the $\sigma$-$\pi$' polarization configuration, the magnetic scattering intensity is proportional to the square of the projection of the magnetic moment, ${\bf M}$, onto the scattered photon wavevector, ${\bf k}_{f}$: $| {\bf M} \cdot {\bf k}_{f}|^{2}$. By rotating the sample by the azimuthal angle $\psi$ (See Fig.~\ref{fig:schematic+temperature}\textbf{a}) with the scattering vector ${\bf Q}={\bf k}_{f}-{\bf k}_{i}$ fixed, the projection $| {\bf M} \cdot {\bf k}_{f}|^{2}$ undergoes modulation as a function of $\psi$. By fitting the $\psi$-dependence of the REXS intensity, the out-of-plane canting angle $\alpha$ (see the inset in Fig.~\ref{fig:azi}\textbf{a}) can be precisely determined.

To obtain the magnetic scattering intensity at each azimuthal angle, we integrated both rocking-curve scans of $\theta$ and $\chi$. One of the difficulties we encountered during our azimuthal dependence study was the mismatch in the beam footprint and the sample shape as $\psi$ varied. To eliminate the magnetic scattering intensity modulation attributed to this beam footprint variation, the magnetic scattering intensity was further normalized with respect to the (0,0,1) structural Bragg peak intensity at each $\psi$. The details of the normalization process are provided in the Supplementary Information.

Fig.~\ref{fig:azi}\textbf{a}-\textbf{b} show the azimuthal dependence of the magnetic scattering intensities at ${\bf Q}_{m1}$ for S1 and ${\bf Q}'_{m2}$ for S3, respectively. Both intensities show local maxima around $\psi$=0$^{\circ}$ and $\psi$=180$^{\circ}$, which suggests that magnetic moments are confined within the ac plane, perpendicular to the zigzag propagation direction, consistent with the previously proposed structure shown in Fig.~\ref{fig:structure+mstructure} \cite{Balz_2021}. The global maximum at $\psi$=180$^{\circ}$ implies that the moment is canted in the positive $\alpha$ direction \cite{Sears2020}. 

The value of canting angle $\alpha$ can be determined by fitting the intensity data. Our best fit result yield $\alpha$=31(2)$^{\circ}$ for S1. However, for S3, determining the azimuthal angle was more challenging due to the weaker signal and the difficulty in normalization as described in Supplementary Information. In order to determine the canted angle in this case, another method was used to support the fitting result. Fig.~\ref{fig:azi}\textbf{c} and \textbf{d} provide a zoomed-in view of $\psi$-range where the magnetic intensity vanishes. The azimuthal angle for zero intensity, $\psi_c$, corresponds to the condition that ${\bf k}_{f}$ is perpendicular to ${\bf M}$ and is very sensitive to the canted angle $\alpha$. For instance, if $\alpha$ changes from 25$^{\circ}$ to 35$^{\circ}$, $\psi_c$ varies from 40$^{\circ}$ to 60$^{\circ}$ for (0,1,1.67), as illustrated in Fig.~\ref{fig:azi}\textbf{d}. The $\psi_c$ could be determined with high precision without the need to normalize intensity between measurements. Both S1 and S3 data give $\alpha=31(2)^{\circ}$ which is consistent with the values obtained using the direct fitting method. Note that S1 and S3 data were obtained for different ${\bf Q}$-vectors, which explains the difference in $\psi_{c}$. For S1, the reference angle 0$^{\circ}$ is along the (1,-3,0) direction. For S3, the reference angle 0$^{\circ}$ is along the (-1,0,0) direction (see Supplementary Information).

\section{Discussion and Conclusions}

We want to point out that the observed canted moment angle of $\alpha=31(2)^{\circ}$ in the current study closely mirrors the result found by Sears et. al.\cite{Sears2020}. This is important since the crystal studied in Ref.~\cite{Sears2020} has a magnetic transition temperature T$_{N}$=12~K, substantially different from the samples studied here. In addition, the magnetic stacking structure of that sample is also different. In Ref.~\cite{Sears2020}, the magnetic Bragg peak was found at (0,1,3/2), indicative of a magnetic structure characterized by a two-layer periodicity, which is a clear departure from the three-layer periodicity found in our study and several other recent neutron scattering studies \cite{sears15,banerjee16,Banerjee2018,Balz_2021}.
The observed robustness of the moment direction $\alpha$, independent of the sample details, is the main result of our study. However, this raises a question as to the origin of the difference in T$_{N}$. 

Variations in interlayer couplings could presumably explain the observed sample dependence of T$_{N}$. The variations can naturally arise from stacking sequence differences. However, the interlayer interactions have been mostly ignored because the magnetic ordering can be well-explained using only the intralayer interactions, and the interlayer interactions are orders of magnitude weaker than their intralayer counterparts \cite{Rau2014,Winter2016,Chaloupka2016,Maksimov2020}. It is interesting to note that stacking structure changes seem to have a dramatic effect on the magnetism of another honeycomb lattice material CrI$_{3}$ \cite{Huang2017,Sivadas2018,Jiang2019,Kong2021}. This material also goes through a structural transition from high-temperature C2/m structure to low-temperature R$\bar{3}$ \cite{McGuire2015}. For this material, each layer orders ferromagnetically with the moment direction pointing perpendicular to the honeycomb planes. However, depending on how the layers are stacked, the system can order antiferromagnetically or ferromagnetically between layers \cite{Sivadas2018,Jiang2019,Kong2021}. It was found that antiferromagnetic order is preferred in the C2/m structure while ferromagnetic order is preferred in the R$\bar{3}$ structure \cite{Sivadas2018}. A recent muon-spin-rotation investigation reported that the magnetic ordering temperature is sensitive to the volume fraction of monoclinic and rhombohedral phases in crystals with coexisting C2/m and R$\bar{3}$ phases \cite{Meseguer2021}. The coexistence of C2/m and R$\bar{3}$ is also frequently observed in $\alpha$-RuCl$_{3}$, especially in thin crystals \cite{Kim2022}, and a similar investigation could be useful to elucidate the origin of the T$_{N}$ variation.

However, the $T_N$ variation in $\alpha$-RuCl$_{3}$ may have a complicated origin, due to the complexity of the in-plane physics in this material. The nearest neighbour J-K-$\Gamma$ model is widely used to describe the magnetism in $\alpha$-RuCl$_{3}$:
\begin{equation}
\label{eqn:jkg_model}
    H=\sum_{<i,j>_{\gamma}} JS_{i}S_{j} + KS_{i}^{\gamma}S_{j}^{\gamma} + \Gamma(S_{i}^{\alpha}S_{j}^{\beta} + S_{i}^{\alpha}S_{j}^{\beta}),
\end{equation}
where $\gamma$ describes the three distinct bonds between nearest neighbour spins $S_{i}$ and $S_{j}$, with mutually exclusive $\alpha, \beta, \gamma \in \{x,y,z\}$. The model consists of an isotropic Heisenberg term (J), a bond-dependent anisotropic Kitaev term (K), and a symmetric off-diagonal ($\Gamma$) term. The presence of relatively large K and $\Gamma$ was found in many works \cite{Rau2014,Winter2016,Chaloupka2016,Chaloupka2015,Rusnacko2019,Lampen-Kelley2018,Sears2020,Suzuki2021,Little2017,Ozel2019,Wu2018,Eichstaedt2019,HSKim2015,Yadav2016,Hou2017,Wang2017,Eichstaedt2019,Suzuki2018,Cookmeyer2018}. The current result again corroborates these earlier studies and confirms the presence of a large $\Gamma$ term. One consequence of large $\Gamma$ is the large magnetic anisotropy between in-plane and out-of-plane susceptibilities \cite{Lampen-Kelley2018,Winter2017,Sears2020}. However, this anisotropy ratio $M_{ab} / M_{c}$ varies from 5 to 8 depending on the sample, strongly suggesting that sample dependence might be present for $\Gamma$ \cite{Kubota2015,Sears2015,weber2016,Banerjee2017,Lampen-Kelley2018,Breitner2023}. While the moment canting angle $\alpha$ is sensitive to the ratio $|\Gamma/K|$, this sensitivity is diminished in the limit of a large $\Gamma$ interaction. According to the classical model \cite{Chaloupka2016}, a wide range of $|\Gamma/K|$ between 1.0 and 1.6 agree within the observed range of $\alpha=31(2)^{\circ}$. Therefore, the variation of $|\Gamma/K|$ in this range would be still compatible with our observation of robustness of $\alpha$.

To summarize, we investigated the sample dependence of the magnetic structure of two $\alpha$-RuCl$_{3}$ samples S1 and S3, characterized by T$_{N}$=7.3~K and T$_{N}$=6.5~K, respectively. Two samples are distinguished clearly in their structures by the domain population. While S1 consists primarily of a single domain of the R$\bar{3}$ structure, S3 has a significant mixing between two twin domains of R$\bar{3}$. However, the magnetic structure of these two samples is consistent with previously determined magnetic structure with three-layer-periodicity \cite{Balz_2021}. Despite the clear difference in $T_{N}$ and domain distribution, we find the canting angle of the ordered moment, $\alpha$, remains unchanged in both samples.

\section{Methods}

All batches of $\alpha$-RuCl$_{3}$ crystals were grown using the chemical vapor transport method described in previous studies \cite{Kim2022}. S1, S2, and S3 were from the same batch while S4 and S5 were grown in separate batches. Two samples, designated as sample 1 (S1) and sample 3 (S3), were selected for single crystal X-ray diffraction measurements performed at the QM2 beamline at Cornell High Energy Synchrotron Source (CHESS). A photon energy of 20keV was employed for studying the samples in transmission geometry. During the data collection, the samples underwent a 360$^{\circ}$ rotation in 0.1$^{\circ}$ steps, and a Pilatus 6M area detector was used to capture the intensity. The temperature was controlled by a helium cryostream onto the $\alpha$-RuCl$_{3}$ crystals, enabling reciprocal maps to be obtained at T=200K and 20K, above and below the structural transition, respectively. 

The same two samples (S1 and S3) were also subjected to resonant elastic X-ray scattering study at the 4-ID beamline at National Synchrotron Light Source II (NSLS II) at Brookhaven National Lab. The incident photon energy was tuned to the ruthenium L$_{3}$ edge (2.837keV). A graphite analyzer (002) was utilized for analyzing the scattered photons, allowing those with polarization parallel to the scattering plane ($\pi$ polarization) to be selected. The samples were cooled using a closed-cycle cryostat, enabling cooling down to T=5K, below the magnetic transition temperature.

\bmhead{Data Availability}
The data that support the findings of this study are available from the corresponding author upon reasonable request.

\bmhead{Code Availability}
The custom codes for analyzing the data and implementing the calculation in this study are available from the corresponding author upon reasonable request.

\bmhead{Acknowledgments}
We thank Jiefu Cen and Hae Young Kee for insightful discussions. Work at the University of Toronto was supported by the Natural Science and Engineering Research Council (NSERC) of Canada, Canadian Foundation for Innovation, and Ontario Research Fund. Resonant elastic x-ray scattering experiment was conducted at 4-ID beamline of the National Synchrotron Light Source II, a U.S. Department of Energy (DOE) Office of Science User Facility operated for the DOE Office of Science by Brookhaven National Laboratory under Contract No. DE-SC0012704. Single crystal X-ray diffraction work is based on research conducted at the Center for High-Energy X-ray Sciences (CHEXS), which is supported by the National Science Foundation (BIO, ENG and MPS Directorates) under award DMR-1829070.

\bmhead{Supplementary Information}
The supplementary material is available at ...

\bmhead{Author Contributions}
The samples were grown by S.K. and prepared together with E.H. for the experiment. The resonant elastic experiments were carried out by S.K. and E.H. with the help of C.N. and the X-ray diffraction experiments were carried out by S.K. and E.H. with the help of J.R. The data was analyzed by S.K.. The idea was conceived by Y.K. and the work was supervised by Y.K. All authors contributed to the writing and preparation of the manuscript.

\bmhead{Competing Interests}
The authors declare no competing interests.

\bibliography{RuCl3refs}
\end{document}


\title{Supplementary Information: Re-investigation of Moment Direction in Kitaev Material $\alpha$-RuCl$_{3}$}

\author[1,2]{\fnm{Subin} \sur{Kim}}\email{bin.kim@mail.utoronto.ca}

\author[1]{\fnm{Ezekiel J.} \sur{Horsley}}\email{ezekiel.horsley@utoronto.ca}

\author[2]{\fnm{Christie S.} \sur{Nelson}}\email{csnelson@bnl.gov}

\author[3]{\fnm{Jacob P. C.} \sur{Ruff}}\email{jpr243@cornell.edu}

\author*[1]{\fnm{Young-June} \sur{Kim}}\email{youngjune.kim@utoronto.ca}

\affil[1]{\orgdiv{Department of Physics}, \orgname{University of Toronto}, \city{Toronto}, \postcode{M5S 1A7}, \state{Ontario}, \country{Canada}}

\affil[2]{\orgdiv{National Synchrotron Light Source II}, \orgname{Brookhaven National Laboratory}, \city{Upton}, \postcode{11973}, \state{NY}, \country{USA}}

\affil[3]{\orgdiv{Cornell High Energy Synchrotron Source}, \orgname{Cornell University}, \city{Ithaca}, \postcode{14853}, \state{NY}, \country{USA}}

\maketitle

\section{Bulk Characterization}\label{Sup1}
\begin{figure} [ht]
    \centering
    \includegraphics[width=0.65\textwidth]{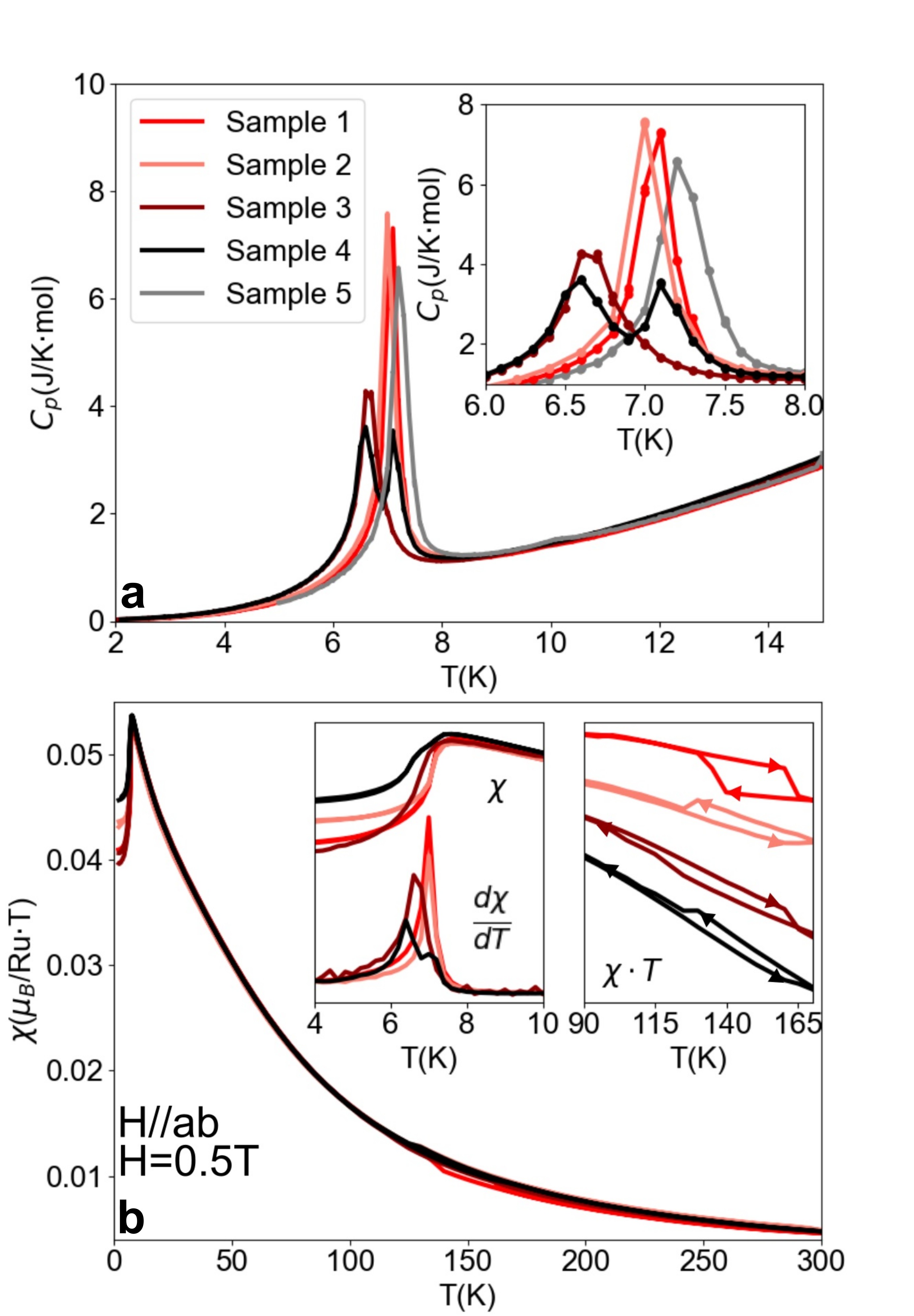}
    \caption{\textbf{Sample dependant specific heat and magnetic susceptibility.} \textbf{a} Temperature dependant specific heat between 5 $\alpha$-RuCl$_{3}$ samples. Peaks are observed only near T$_{N}$=7~K with no anomaly between 10~K to 14K between S1-4 while a minor anomaly is visible for S5.  Inset of \textbf{a} zooms in close to T$_{N}$=7~K. The peak locations vary from T$_{N}$=6.5~K to T$_{N}$=7.3~K between samples with S4 showing two peaks at T$_{N}$=6.5~K and T$_{N}$=7.3~K. \textbf{b} Sample-dependant magnetic susceptibility as a function of temperature with 0.5~T field applied within the honeycomb plane. The measurements were performed for both field cooling and zero-field cooling direction. The first inset of \textbf{b} shows magnetic susceptibility ($\chi$) and its derivative (d$\chi$/dT) close to the magnetic transition temperature T$_{N}\sim$7~K. The sharp drop in susceptibility (or peak in d$\chi$/dT) is consistent with the peak position in specific heat shown in panel \textbf{a}. The second inset of \textbf{b} shows magnetic susceptibility multiplied with temperature. The arrow indicates the measurement direction as well as where the structural transition occurs.}
    \label{fig:characterization}
\end{figure}

\subsection{Specific Heat}

Fig.~\ref{fig:characterization}\textbf{a} displays the temperature-dependent specific heat of the five $\alpha$-RuCl$_{3}$ crystals near the magnetic transition. Distinct peaks are clearly observed around 7~K, indicating magnetic transitions. No additional peaks are discernible between 10~K and 14~K for samples S1 to S4, which rules out the presence of additional magnetic transitions commonly associated with samples containing many stacking faults [1]. However, for S5, a small bump appears at T$_{N}$=10~K, indicating the existence of some stacking faults within the sample. 

A notable difference in peak positions is evident among the samples with N\'eel temperature varying from T$_{N}$=6.5~K to T$_{N}$=7.3~K, as shown in the inset of Fig.~\ref{fig:characterization}\textbf{a}. Furthermore, S4 exhibits two distinct transitions at T$_{N}$=6.5~K and T$_{N}$=7.3~K, illustrating the coexistence of these two phases. A qualitative difference is observable in the peak shapes between the transitions around T$_{N}$=7.3~K (S1, S2, S5) and T$_{N}$=6.5~K (S3, S4). The peak height is approximately twice as large for the T$_{N}$=7.3~K samples compared to the T$_{N}$=6.5~K samples. The peak width is broader for T$_{N}$=6.5~K samples, and exhibits a more pronounced low-temperature tail compared to the T$_{N}$=7.3~K samples. However, we found that the magnetic entropy change across the magnetic transition is similar in these samples. This is estimated by integrating C(T)/T after subtracting the phonon contribution between T$_{N}$=7.1~K to 7.3~K and T$_{N}$=6.4~K to 6.6~K, and we found the values to be 0.7 J/(K$\cdot$mol) in both types of samples, in good agreement with the findings of Widmann et al [2]. 

\subsection{Magnetic Susceptibility}

Temperature-dependent magnetic susceptibility was measured for these samples, as shown in Fig.~\ref{fig:characterization}\textbf{b}. The measurements were conducted with the magnetic field applied along an unspecified in-plane direction for each sample, both in the field-cooling and zero-field cooling conditions. All samples exhibit a drop in magnetic susceptibility below T$_{N}$=7~K, as expected for an antiferromagnetic transition. 

For a closer examination of this transition, the left inset of Fig.~\ref{fig:characterization}\textbf{b} illustrates the magnetic susceptibility near the magnetic transition temperature, along with its first derivative, $d\chi/dT$. The temperature at which the susceptibility drops, equivalent to the peak position in $d\chi/dT$, varies among the samples. The observed peak positions in $d\chi/dT$ align well with the peak positions observed in the specific heat, providing further confirmation of the sample-dependent transition temperature. A qualitative difference is also evident in the magnetic susceptibility behavior between the magnetic transitions at T$_{N}$=7.3~K and T$_{N}$=6.5~K. The drop is much sharper in the case of T$_{N}$=7.3~K samples, as indicated by the pronounced peak in $d\chi/dT$.

Additionally, all samples undergo structural transitions around T$_{s}\approx$150~K. The right inset of Fig.~\ref{fig:characterization}\textbf{b} depicts the magnetic susceptibility multiplied by temperature in the vicinity of the structural transition temperature. The arrow indicates the direction of temperature change. Notably, all samples display hysteresis behavior in susceptibility between cooling and heating, indicating the first-order nature of the structural transition. Note that the field direction was not the same for all samples measured. This explains the difference in the thermal hysteresis behavior. For example, in S1, the susceptibility drops above the structural transition temperature, while in S2, the susceptibility increases. However, the magnetic susceptibilities converge and become consistent across all samples below the structural transition temperature.

In the case of S1 and S2, the hysteresis ranges are approximately 20~K and 30~K, respectively, with abrupt susceptibility jumps across the structural transition. However, for S3, the hysteresis range spans around 70~K, exhibiting a more gradual susceptibility change upon cooling. Similarly, S4 seems to exhibit two types of structural transitions: one with a sharp drop around 130~K (40~K hysteresis range), akin to S1 and S2, and another with a gradual susceptibility shift spanning about 70~K. The presence of two types of structural transitions in S4 indicates the coexistence of T$_{N}$=7.3~K and T$_{N}$=6.5~K samples. This observation suggests that samples with lower T$_{N}$=6.5~K tend to have larger hysteresis ranges compared to those with T$_{N}$=7.3~K, consistent with previous report [3]. However, it is worth noting that the hysteresis range can vary from 20~K to 90~K even for samples with the same T$_{N}$. 

\section{Azimuthal Angle Reference}\label{Sup2}

\begin{figure} [ht]
    \centering
    \includegraphics[width=\textwidth]{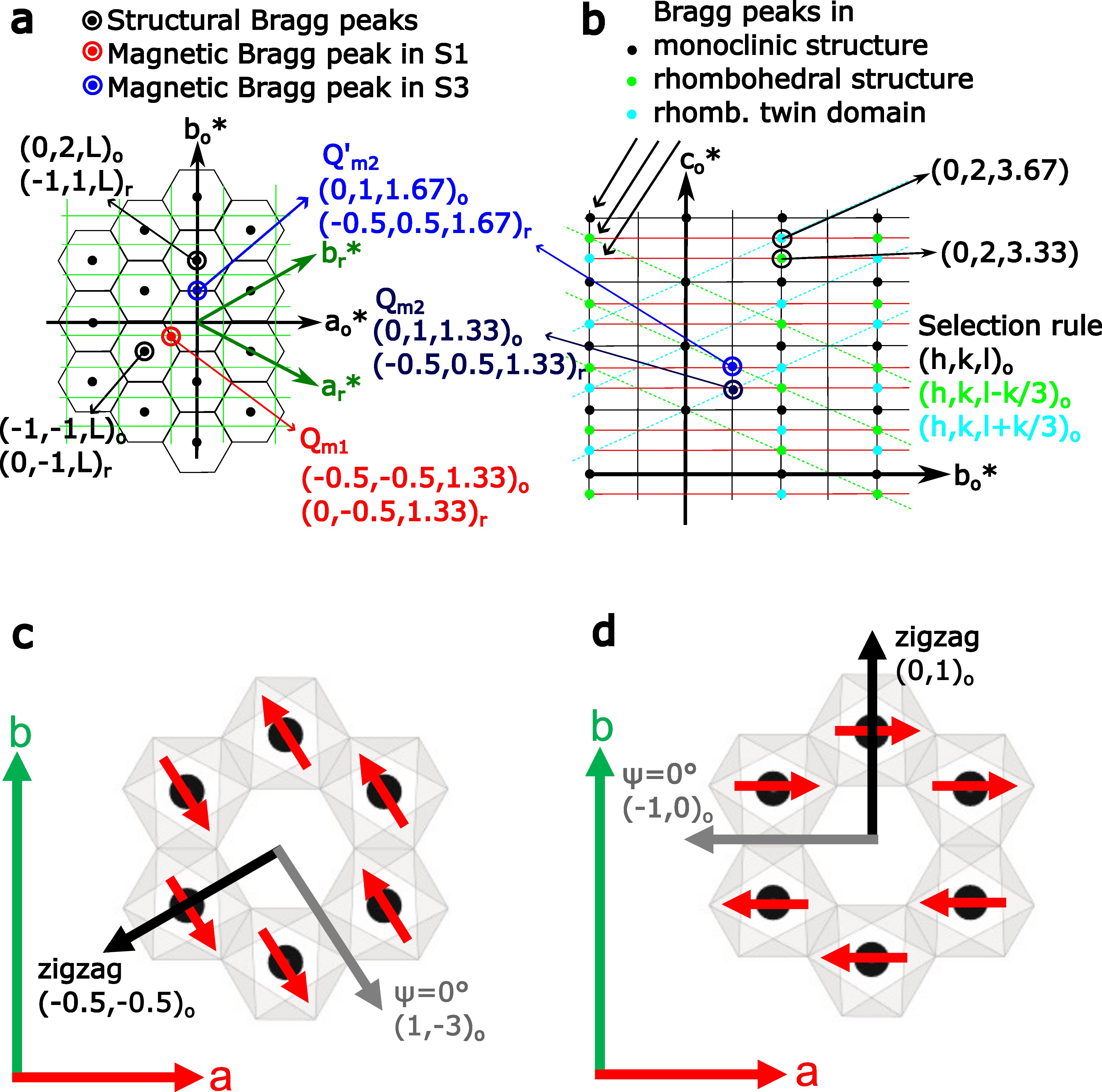}
    \caption{\textbf{Pseudo-orthorhombic notation and azimuthal angle reference.} \textbf{a}-\textbf{b} Illustrations of the pseudo-orthorhombic notation. \textbf{a} Comparison of the orthorhombic notation with rhombohedral notation in the a$^{*}$-b$^{*}$ plane. \textbf{b} Comparison of the orthorhombic with other coordinate systems b$^{*}$-c$^{*}$ plane. Allowed Bragg peak positions for different structures are indicated by coloured dots on the reciprocal plane. \textbf{c}-\textbf{d} Illustrations of the magnetic structure for \textbf{a} S1 and \textbf{b} S3 respectively. A black arrow indicates the zigzag propagation direction and a corresponding grey arrow indicates the reference vector respect to the zigzag direction. }
    \label{fig:azimuth_ref}
\end{figure}

The reference angle ($\psi$=0$^{\circ}$) was chosen differently for S1 and S3 due to the locations of magnetic Bragg peaks. In both samples, the reference angle was chosen such that it corresponds to the direction perpendicular to the zigzag propagation. For S1, the magnetic peak was observed at (-0.5,-0.5,L), and the reference angle was chosen to be (1,-3,0) (see Fig.~\ref{fig:azimuth_ref}\textbf{c}). For S3, the magnetic peak was observed at (0,1,L), and the reference angle was chosen to be (-1,0,0) instead (see Fig.~\ref{fig:azimuth_ref}\textbf{d}).

\section{Data Analysis of the Azimuthal Dependence}\label{Sup3}

\begin{figure} [ht]
    \centering
    \includegraphics[width=0.8\textwidth]{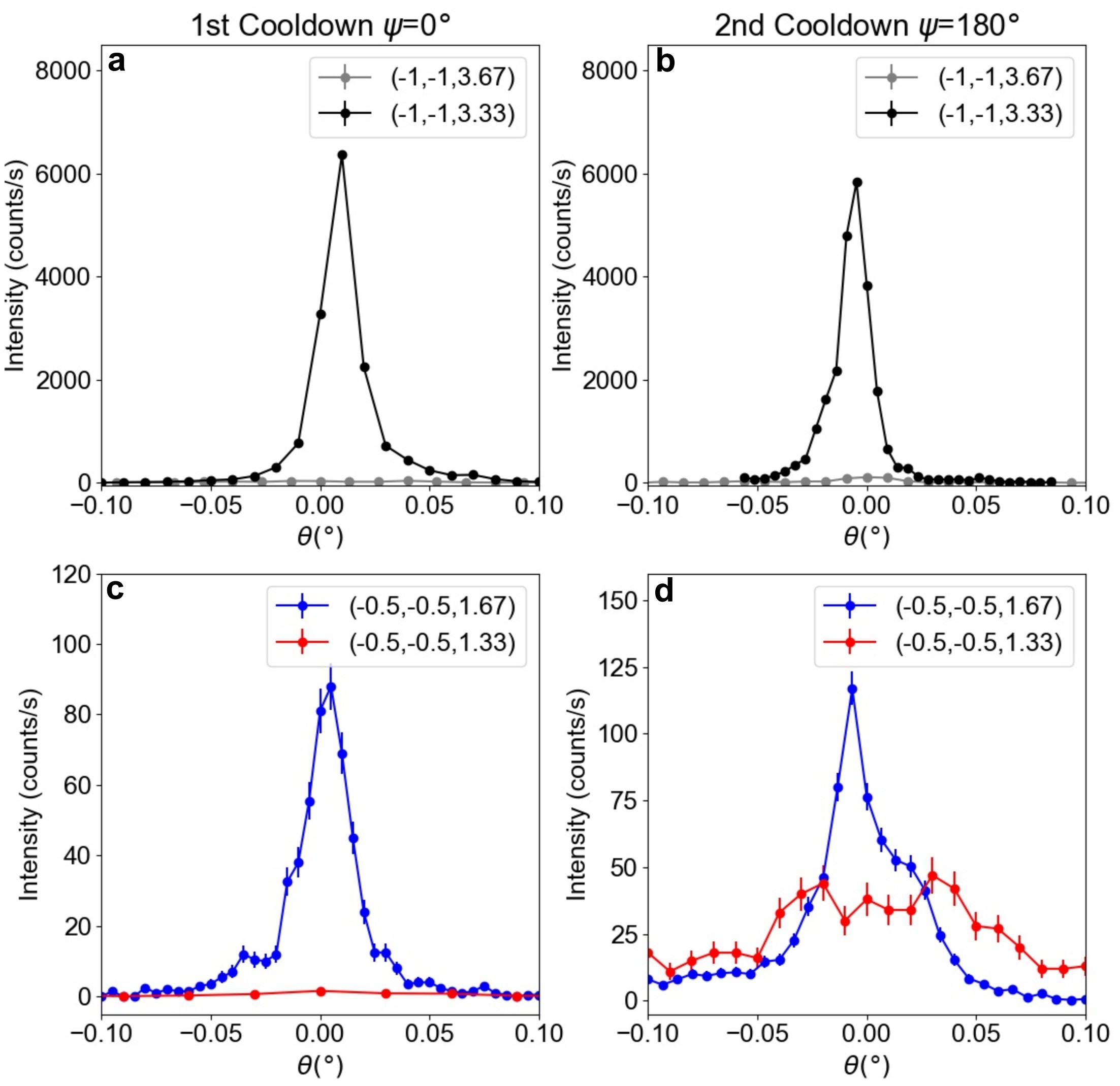}
    \caption{\textbf{Structural and magnetic domain comparison between two thermal cycles.} The rocking curves of two structural Bragg peaks; ${\bf Q}^{\prime}_{s1}$=(-1,-1,3.33) and ${\bf Q}_{s1}$=(-1,-1,3.67), of S1 in the first cooling cycle \textbf{a} and in the second cooling cycle \textbf{b} at different azimuthal angles. The intensity of ${\bf Q}^{\prime}_{s1}$ dominates in both temperature cycles and the structure of S1 consists primarily of a single domain of R$\bar{3}$. The rocking curves of two magnetic Bragg peaks; ${\bf Q}_{m1}$=(-0.5,-0.5,1.33) and ${\bf Q}^{\prime}_{m1}$=(-0.5,-0.5,1.67), of S1 in the first cooling cycle \textbf{c} and in the second cooling cycle \textbf{d}. A clear change in the intensities ratio is observed between two magnetic Bragg peaks across the temperature cycles. The change in the intensity ratio is attributed to the difference in the magnetic domain distribution after the thermal cycling.}
    \label{fig:Temp_Cycle}
\end{figure}

\subsection{Normalization of Two Separate Measurements}

The azimuthal dependence was carried out over two separate beamtimes. This was necessary because, for each measurement, the sample could only rotate about $\psi$ $\approx$ 90$^{\circ}$ to 100$^{\circ}$ due to instrumental limitations. Therefore, it is unlikely for the same part of a sample to be probed in these two beamtimes. Additionally, there is a possibility of a magnetic domain change due to thermal cycling (see Fig.~\ref{fig:Temp_Cycle}). These factors can potentially affect the relative intensity between two measurements, necessitating the need to normalize the intensity. The normalization of two separate measurements was achieved by matching the overlapping regions.

\begin{figure} [ht]
    \centering
    \includegraphics[width=1\textwidth]{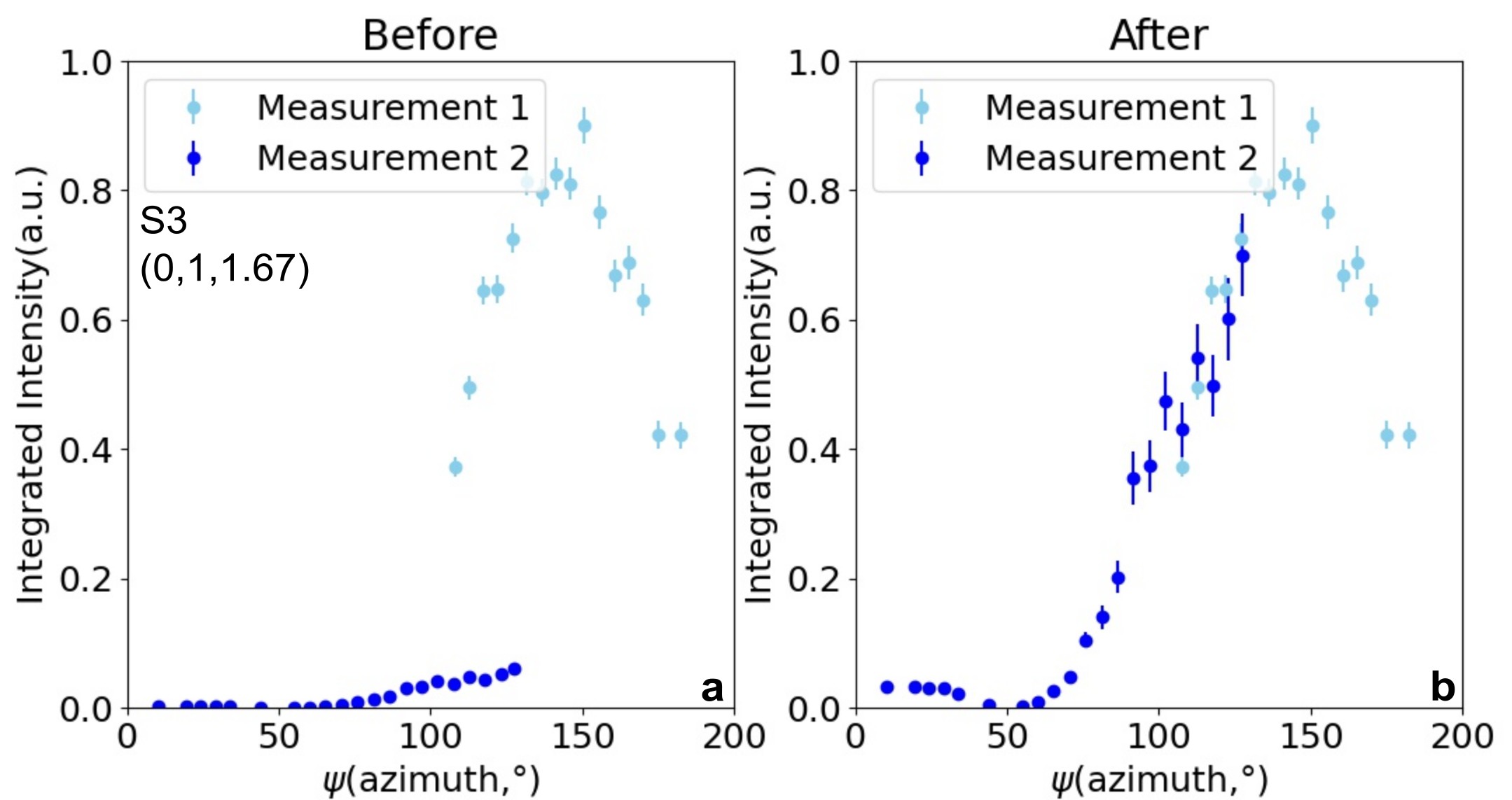}
    \caption{\textbf{Normalization process of two separate measurements in S3.} \textbf{a} Direct comparison of magnetic intensity of ${\bf Q}^{\prime}_{m2}$=(0,1,1.67) as a function of $\psi$ in S3. The dark and light blue data indicate two separate measurements that had to be performed to cover about 200$^{\circ}$ in $\psi$. \textbf{b} Azimuthal dependence of magnetic intensity after normalization. The intensity was normalized by scaling the intensities of two measurements to agree in the overlapping region of $\psi$.}
    \label{fig:Raw data S3}
\end{figure}

As an example, in Fig.~\ref{fig:Raw data S3}\textbf{a}, an azimuthal dependence in magnetic intensity of ${\bf Q}^{\prime}_{m2}$=(0,1,1.67) is shown, where the intensity of two separate measurements are directly compared. A disagreement in intensities was observed in these measurements between $\psi$=100$^{\circ}$ and $\psi$=130$^{\circ}$. Therefore, two measurements were normalized by scaling the two intensities to agree and the result is shown in Fig.~\ref{fig:Raw data S3}\textbf{b}. The significant discrepancy between these measurements may have occurred due to a change in the magnetic domain population following the temperature cycle.

\subsection{Removal of Beam Footprint Contribution}

\begin{figure} [ht]
    \centering
    \includegraphics[width=0.8\textwidth]{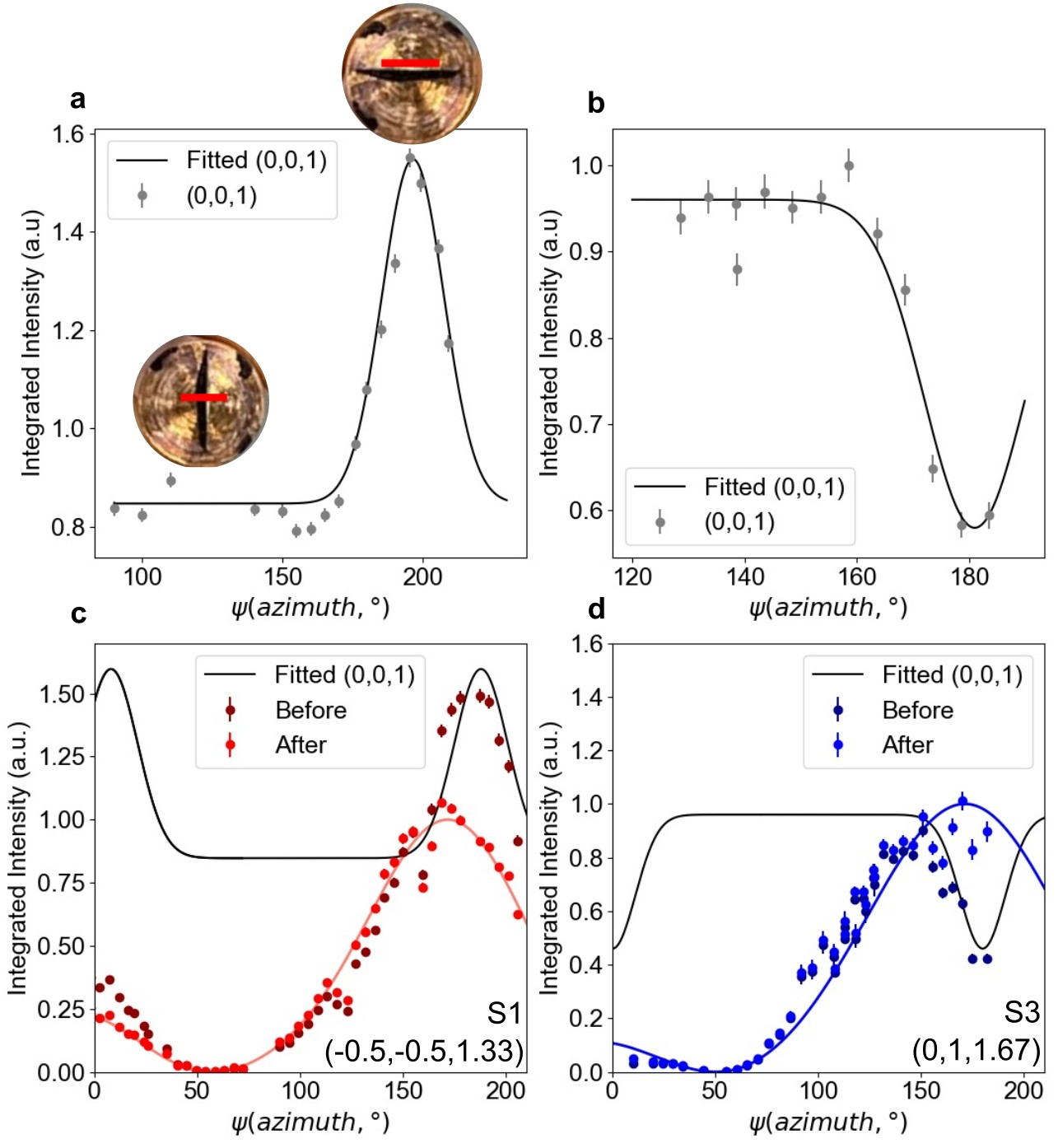}
    \caption{\textbf{Removal of beam-footprint modulation in the magnetic intensity.} \textbf{a}-\textbf{b} Azimuthal dependence of the structural Bragg peak (0,0,1) in S1 and S3, respectively. The modulation occurred due to the change in the area covered by the beam on the sample which is illustrated as two pictures in \textbf{a} where the red line indicates the X-ray beam covers the black $\alpha$-RuCl$_{3}$ crystal differently at different $\psi$. Smooth lines were fitted to empirical data and were used to account for the beam footprint modulations in the magnetic intensities. \textbf{c}-\textbf{d} Removing the modulation contribution from the beam footprint in the azimuthal dependence of the magnetic intensity in S1 and S3, respectively. The dark red (dark blue) data represents the results obtained after normalizing two separate measurements. The black lines represent the fitted modulations of the structural Bragg peak intensity. The light red (blue) lines represent the intensity modulation of the magnetic intensity after normalizing with respect to the structural Bragg peak. This was provided in the main text as Fig. 5\textbf{a}-\textbf{b}}
    \label{fig:beam_footprint}
\end{figure}

A modulation in structural intensity of (0,0,1) was observed in both samples which are shown in Fig.~\ref{fig:beam_footprint}\textbf{a}-\textbf{b}. In the ideal case, where the size of the X-ray beam is smaller than the sample, no modulation in intensity is expected. However, when the beam size exceeds that of the sample, modulation occurs due to the change in the area covered by the beam on the sample. This is illustrated in inset images in Fig.~\ref{fig:beam_footprint}\textbf{a}. 

Therefore, it is necessary to remove the contribution from the beam footprint modulation in the azimuthal dependence of the magnetic intensity. Fig.~\ref{fig:beam_footprint}\textbf{c}-\textbf{d} illustrates the process of removing the beam footprint contribution by normalizing the magnetic intensity with respect to the structural (0,0,1) Bragg peak. The normalized data are present in the main paper as the finalized data (Fig. 5\textbf{a}-\textbf{b}).

\section{Azimuthal Angle at the Vanishing Magnetic Intensity}\label{Sup4}

\begin{figure} [ht]
    \centering
    \includegraphics[width=0.9\textwidth]{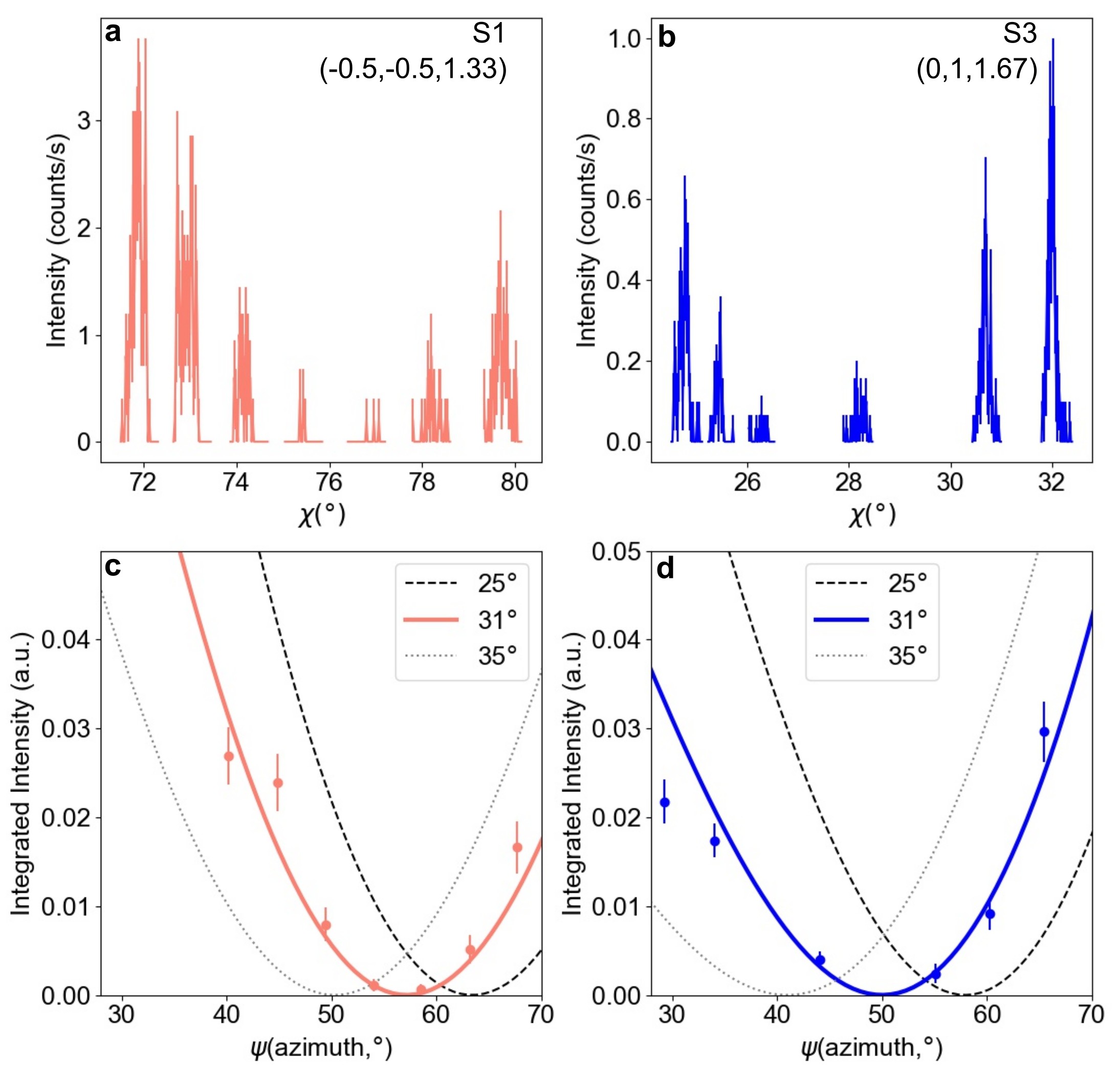}
    \caption{\textbf{Azimuthal dependence of magnetic intensities at the vanishing angle $\psi_{c}$.} \textbf{a}-\textbf{b} Rocking curves of the magnetic intensity measured at various azimuthal angles close to the region where the magnetic intensity vanishes. The data were collected by counting 20 seconds per data point to accurately capture the modulation of the magnetic intensity. \textbf{c}-\textbf{d} Azimuthal dependence of the magnetic intensities close to $\psi_{c}$ in S1 and S3, respectively. The lines indicate the simulated data with canted moment angles $\alpha$=35$^{\circ}$,31$^{\circ}$, and 25$^{\circ}$ which are ordered from left to right. This was shown in Fig. 5\textbf{c}-\textbf{d} in the main text.}
    \label{fig:SFig6}
\end{figure}

In the main text, we obtained the moment canting angle by fitting the overall azimuthal angle dependence. However, one caveat of this fitting procedure is its reliance on the normalization discussed in Section~\ref{Sup3}. To address this concern, we employed an alternative method to determine the magnetic moment direction without the need for a normalization process. This is achieved by measuring the azimuthal angle $\psi_{c}$ at which the magnetic intensity vanishes.

Fig.~\ref{fig:SFig6}\textbf{c}-\textbf{d} displays the azimuthal dependence of the magnetic intensity around the $\psi_{c}$ at which the magnetic intensity vanishes. The simulated magnetic intensities, each calculated with a canted moment angle of $\alpha=25^{\circ}$, $31^{\circ}$, and $35^{\circ}$, demonstrate a clear dependence on $\psi_{c}$ with respect to the canted angle $\alpha$. The observed vanishing angle agrees the best with the calculation using $\alpha=31^{\circ}$ in both samples, confirming that the magnetic moment directions are identical between the two samples as discussed in the main text.
